\pdfoutput=1
\documentclass[aps,prb,twocolumn,groupedaddress]{revtex4}

\usepackage{xspace}
\usepackage{graphicx}
\usepackage{epsfig}
\usepackage{amsfonts}
\usepackage{amssymb}
\usepackage[usenames]{color}

\begin{document}

%\preprint{}

\title{\textsf{Polarization-Induced Zener Tunnel Junctions in Wide-Bandgap Heterostructures}}

\author{John Simon}
\email[Electronic mail: ]{jsimon@nd.edu}
\author{Ze Zhang}
\author{Kevin Goodman}
\author{Thomas Kosel}
\author{Patrick Fay}
\author{Debdeep Jena}
\affiliation{Department of Electrical Engineering, University of Notre Dame, Notre Dame, IN 46556, USA}
\date{\today}

\begin{abstract}
The large electronic polarization in III-V nitrides allow for novel physics not possible in other semiconductor families.  In this work, interband Zener tunneling in wide-bandgap GaN heterojunctions is demonstrated by using polarization-induced electric fields.  The resulting tunnel diodes are more conductive under reverse bias, which has applications for zero-bias rectification and mm-wave imaging.  Since interband tunneling is traditionally prohibitive in wide-bandgap semiconductors, these polarization-induced structures and their variants can enable  a number of devices such as multijunction solar cells that can operate under elevated temperatures and high fields.

\end{abstract}

%\pacs{81.10.Bk, 72.80.Ey} \keywords{***}

\maketitle
%%%%% Text of paper:
%%

%=================================================================
%INTRODUCTION & MOTIVATION

{\em Introduction:} Since Esaki's explanation of the interband electron tunneling process\cite{Esaki1} across heavily doped Germanium p-n junction diodes in 1957, the phenomena has been investigated in great detail in many semiconductors\cite{sze}.  Due to negative differential resistance, negligible minority carrier storage and the resulting high speeds, tunnel diodes have found usage in applications such as oscillators\cite{Persson}, amplifiers\cite{Chang1}, frequency converters\cite{Sterzer}, and detectors\cite{Montgomery}.  A notable and increasingly important application of tunnel junctions is in electrically connecting semiconductor p-n junctions of disparate bandgaps, such as in multijunction solar cells \cite{Miller,DeSalvo}. 

Interband tunneling of electrons in semiconductors is impeded by two factors: the tunneling barrier height determined by the bandgap, and the tunneling barrier thickness.  For wide bandgap semiconductors such as the III-V nitrides (GaN, AlN) and SiC, tunneling is low due to the high barrier heights, and is hampered further by the inability to achieve degenerate n- and p-type impurity doping.  However, by utilizing the giant built-in electronic polarization fields present in wurzite III-V nitride semiconductor heterostructures\cite{Bernardini, djPolBook}, it is possible to achieve interband tunneling in p-n junction diodes.  This principle was recently used for connecting two nitride p-n junctions to demonstrate a multicolor light emitter \cite{Grundmann}.  Such polarization-induced tunnel junctions can enable Ohmic contacts to the valence band of large bandgap semiconductors.  They will enable multicolor light emitters/detectors and multijunction solar cells by taking advantage of the large span in direct bandgaps (IR to deep- UV) covered by the III-V nitride semiconductor family.  Wide bandgap semiconductors also exhibit long lifetimes for spin-polarized carriers and room-temperature ferromagnetic behavior; tunnel junctions can enable efficient injection of polarized carriers for spintronic devices.  In light of the importance, this work presents an experimental demonstration and a theoretical model for polarization-induced tunnel junctions in III-V nitride heterostructures.  GaN-AlN-GaN p-n tunnel junctions are shown to operate as backward diodes for zero-bias detectors at room temperature.

%=================================================================
\begin{figure}
\includegraphics[width=3.3in]{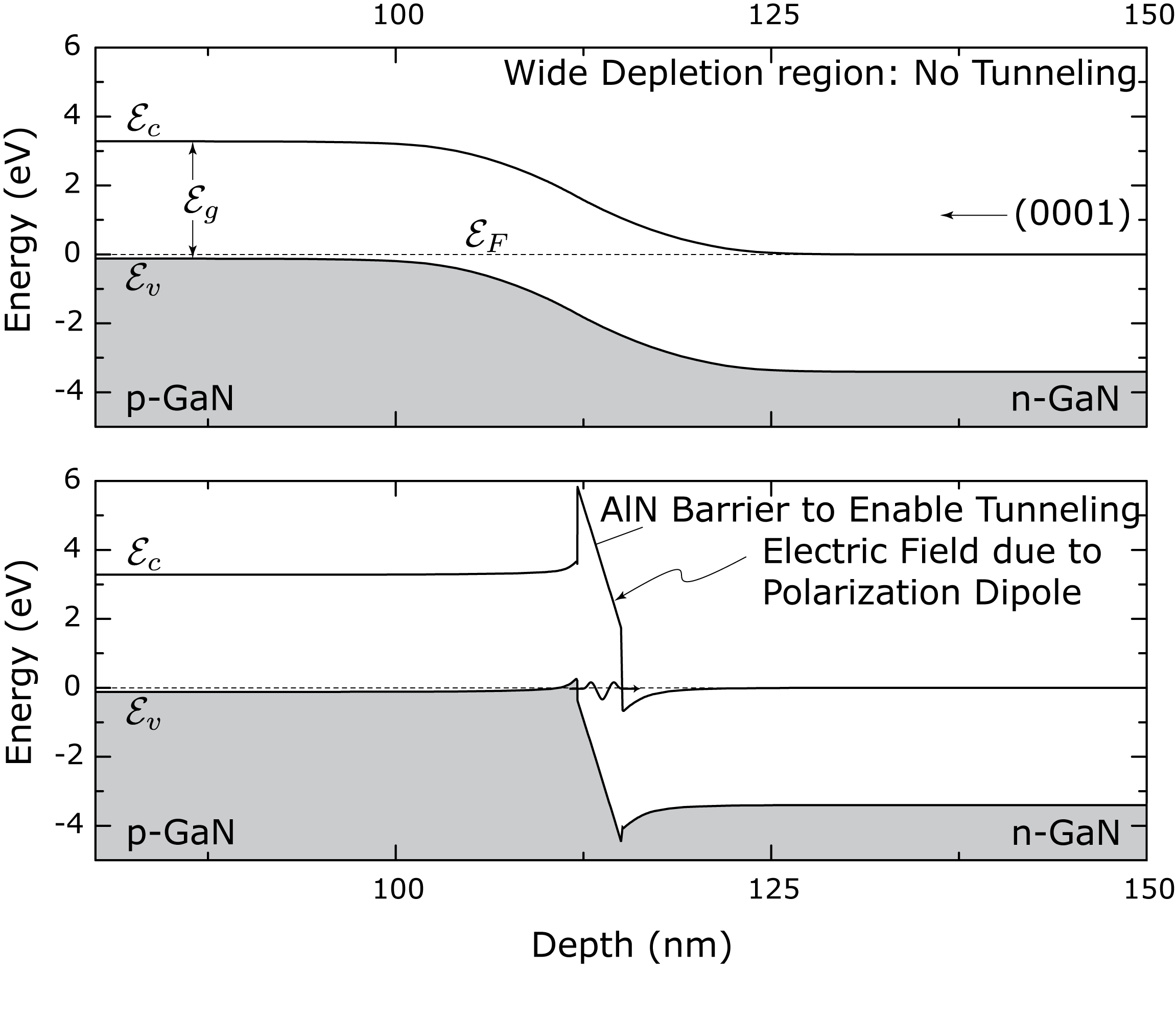}\\
\caption{Energy band diagrams of a normal p-n junction and a polarization-induced tunneling p-n junction.  The polarization-induced electric field shrinks the depletion region and aligns the valence and conduction bands to facilitate interband tunneling.}
\label{fig1}
\vspace{-4ex}
\end{figure}
%=================================================================

Wurtzite III-V nitride semiconductors (GaN, AlN) exhibit giant polarization fields in the 0001 crystal direction \cite{Bernardini, djPolBook}. When a thin strained AlN layer is inserted into a GaN crystal along the 0001 direction, the abrupt discontinuity of spontaneous polarization and the piezoelectric strain leads to the formation of a dipole of fixed sheet charges at the two AlN/GaN heterojunctions.  This polarization sheet charge density is as large as $\sigma_{\pi} \sim 6 \times 10^{13}$/cm$^{2}$.  In the absence of other charges, the electric field inside the AlN layer as a result is $\mathcal{F}_{\pi} = q \sigma_{\pi}/ \epsilon_{s} \sim 12 $ MV/cm, where $q$ is the electron charge and $\epsilon_{s} \sim 9 \epsilon_{0}$ is the dielectric constant of AlN.  This is an extremely high field; it cannot be attained by conventional impurity doping, and is the critical enabler of interband (Zener) tunneling.  For comparison, Figure \ref{fig1} shows the energy band diagram calculated for an impurity-doped GaN:Mg/GaN:Si p-n junction heavily doped at $N_{A} \sim 10^{19}$/cm$^{3}$ and $N_{D} \sim 9 \times 10^{18}$/cm$^{3}$ respectively.  The depletion region width (effective tunneling barrier thickness) is $W \sim 25$ nm, and the maximum electric field reached at the junction is $\mathcal{F}_{max} \sim 2 \mathcal{E}_{g}/W \sim 2.7 $ MV/cm, where $\mathcal{E}_{g} \sim 3.4 $ eV is the bandgap of GaN.  The same figure also shows the energy band diagram calculated by a self-consistent Schr\"{o}dinger-Poisson solution for the case when a $t_{AlN} = 3$ nm AlN tunnel barrier is inserted at the p-n junction.  The polarization dipole shrinks the depletion width to a thickness close to $t_{AlN}$ and the band bending of the order of $\sim q \mathcal{F}_{\pi} t_{AlN}$ realigns the band edges.  Interband tunneling of electrons from the valence band (VB) in the p-type side to the conduction band (CB) in the n-type side is then possible under reverse bias voltage; such polarization-induced tunnel junctions thereby conduct more current under reverse bias, and are `backward' diodes\cite{sze}.  

%This optimum AlN thickness has to be experimentally determined since values of the polarization charges determined above are estimates based on the %results by Bernardini\cite{Bernardini} and have yet to be experimentally verified.

\begin{figure}
\includegraphics[width=3.3in]{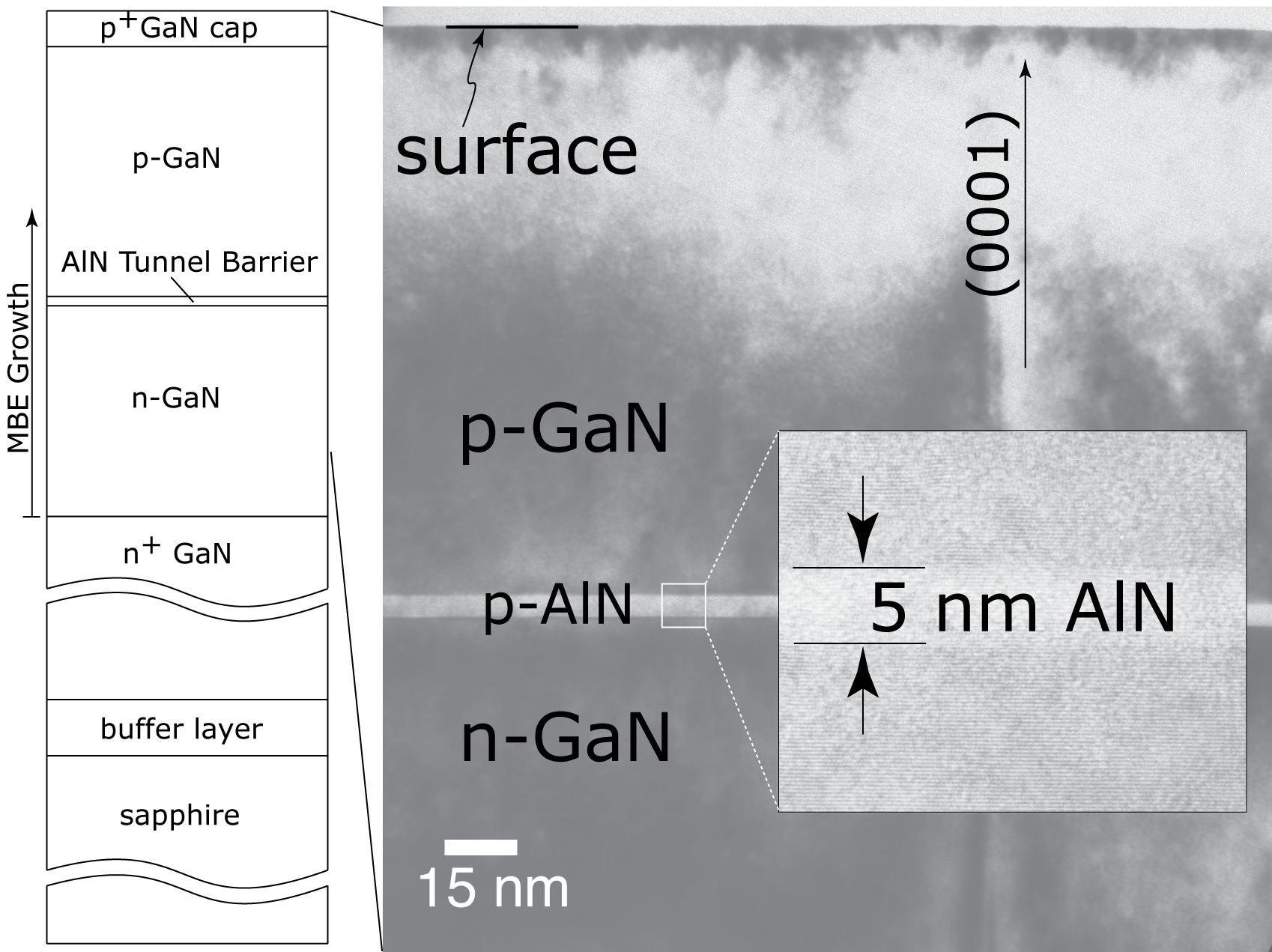}\\
\caption{Layer structure and a TEM cross section image of the 5 nm AlN backward diode.  The insert shows a high resolution image of the AlN layer verifying the AlN thickness.}
\label{fig2}
\vspace{-4ex}
\end{figure}

%=================================================================
%EXPERIMENT
{\em Experiment:} In order to investigate this phenomena, GaN-AlN-GaN heterostructures were grown by plasma-assisted Molecular Beam Epitaxy (MBE) on commercially available n-type doped GaN 0001 templates.  The growth was performed at a substrate thermocouple temperature of 600$^{\circ}$C and a RF plasma power of 275 W, which corresponds to a growth rate of $\sim$ 150 nm/hr.  A 100 nm Si-doped GaN layer ($N_{D} \sim 9 \times 10^{18}$/cm$^{3}$) was grown, followed by a Mg doped AlN layer of thickness $t_{AlN}$, and a 100 nm p-type GaN with Mg doping $N_{A} \sim 10^{19}$/cm$^{3}$.  The thickness of the AlN tunnel-barrier was varied over 6 samples ($t_{AlN}$ = 0, 1.4, 2.8, 3.5, 4.3 \& 5 nm).  The samples were then capped with a heavily p-type doped GaN layer for improved Ohmic contacts.  Fig \ref{fig2} shows the layer structure, and a Transmission Electron Microscope (TEM) image of the GaN/AlN/GaN tunnel junction which was used to confirm the thickness and uniformity of the AlN layer.  The TEM image also indicates a sharp heterojunction with high crystalline quality as inferred from lattice imaging.

%=================================================================
%=================================================================

\begin{figure}
\includegraphics[width=3.3in]{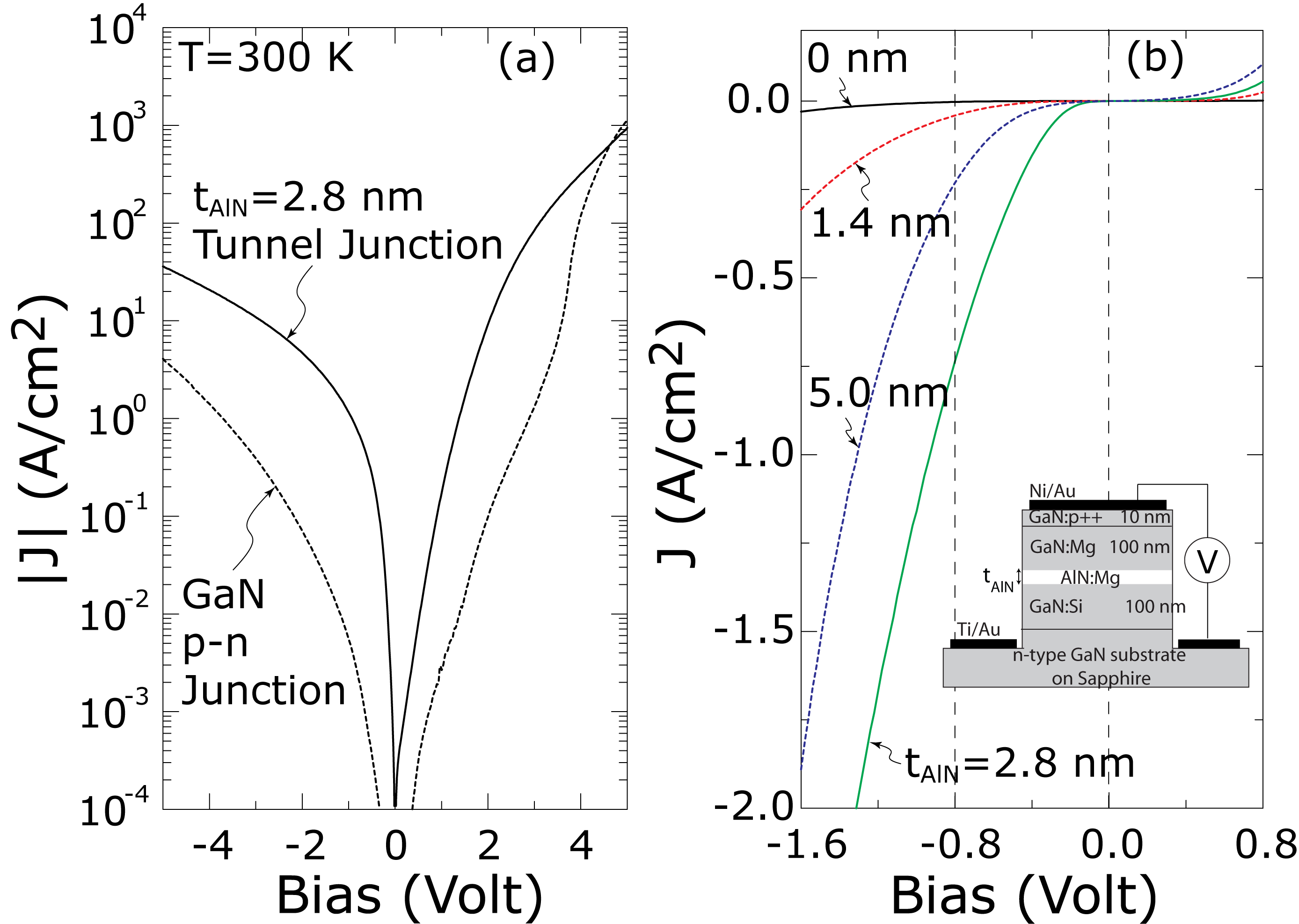}\\
\caption{(a) Measured current-voltage characteristics for a normal  ($t_{AlN}$=0 nm) and a tunnel p-n junction ($t_{AlN}$= 2.8  nm).  (b) Expanded current voltage characteristics under small applied biases showing backward diode behavior.  The 2.8 nm AlN structure exhibits the highest reverse bias tunneling current, as explained using a model that includes polarization-induced electric field.  The inset shows the device structure.}
\label{fig3}
\vspace{-4ex}
\end{figure}

Following MBE growth, the samples were processed into p-n junctions by etching down to the n-type substrate using Cl$_{2}$ plasma in an Inductively Coupled Plasma (ICP) etcher.  Ni/Au and Ti/Au Ohmic contacts to the p and n-type layers respectively were deposited in an electron beam evaporator.  A schematic of the finished device is shown as an inset in Fig \ref{fig3}(b).  Current-voltage characteristics of the junctions were measured at 300 K in a probe station using a semiconductor parameter analyzer.  The results are shown in figure \ref{fig3}.  The devices with the AlN interlayer show clear backward-diode behavior: they conduct more current when reverse biased than when forward biased.  This is a signature of interband tunneling.  An useful metric used for backward diodes is the curvature of the I-V curve at zero voltage bias, defined as $\gamma = (\partial^{2}I/ \partial V^{2}) / (\partial I/ \partial V)$.  The highest $\gamma = 21$ V$^{-1}$ was obtained for a barrier thickness $t_{AlN} = 2.8$ nm.  As indicated in Fig \ref{fig3}, the tunnel junction with $t_{AlN} = 2.8$ nm showed the highest tunneling current density, reaching levels of $\sim$ 750 mA/cm$^2$ at a reverse bias of -0.8 Volt.

%=================================================================
%RESULTS
%=================================================================
%=================================================================

%=================================================================
%\begin{figure}
  %\includegraphics[width=3.0in]{figures/Model}\\
  %\caption{Banddiagram schematic under -0.5 V reverse bias for GaN-AlN-GaN p-n junctions. The potential energy (P.E.(x)) observed by an electron tunneling from the valence band of the p-type region to the conduction band of the n-type region is sketched on the right hand side. }\label{fig:model}
%\end{figure}
%=================================================================

\begin{figure}
\includegraphics[width=3.3in]{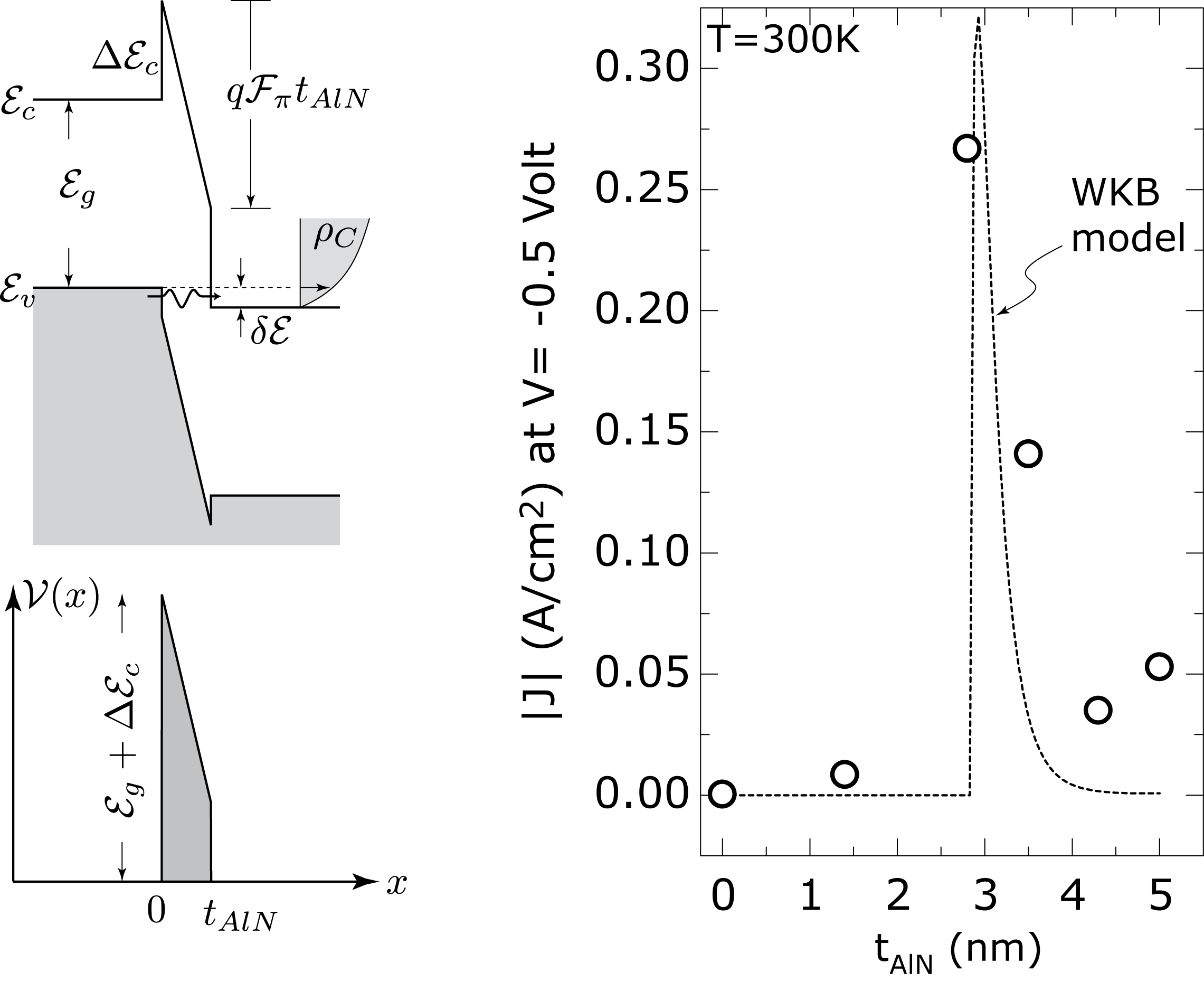}\\
\caption{Band diagram, tunneling barrier, and calculated + measured dependence of tunneling current on AlN thickness (t$_{AlN}$) measured at -0.5 V bias. The experimental results agree with the model of equation \ref{jtunneldep}.}
\label{fig4}
\vspace{-4ex}
\end{figure}

At a fixed reverse bias voltage, the current density, when plotted for various AlN layer thicknesses ($t_{AlN}$) showed a clear maximum.  Figure \ref{fig4} shows the tunnel current densities (circles) at a bias of -0.5 Volt as a function of the AlN layer thickness.  This behavior can be explained using a simple model of interband tunneling coupled with the high polarization fields that enable it.  An energy band diagram as sketched in Figure \ref{fig4} shows the band alignments at a small reverse bias, assuming close to flat-band conditions in the p- and n-type regions.  This form of the band diagram is confirmed by self-consistent Schr\"{o}dinger-Poisson simulations.  For electrons in the valence band of the p-type GaN region to tunnel into the conduction band of the n-type region, the CB edge $\mathcal{E}_{c}$ in the n-side must be lower in energy than the VB edge $\mathcal{E}_{v}$ in the p-side.  As can be seen in the energy band diagram, at an applied reverse bias of $qV$, energy conservation requires 
\begin{equation}
\mathcal{E}_{g} + \Delta \mathcal{E}_{c} =  q \mathcal{F}_{\pi} t_{AlN} + \Delta \mathcal{E}_{c} - \delta \mathcal{E},
\end{equation} 
where $\Delta \mathcal{E}_{c} \sim 2.1 $ eV is the conduction band offset between GaN and AlN, $\mathcal{F}_{\pi} \sim 12$ MV/cm is the polarization-induced electric field, and $\delta \mathcal{E} = qV + q \mathcal{F}_{\pi} t_{AlN} - \mathcal{E}_{g}$.  The critical AlN thickness for which the onset of interband tunneling occurs for the smallest applied voltages is given by $\delta \mathcal{E} \approx 0$, or $t_{AlN}^{cr} \approx \mathcal{E}_{g} / q \mathcal{F}_{\pi}$, which evaluates to $t_{AlN}^{cr} \sim 2.8$ nm, essentially identical to the diode that exhibits the highest tunneling current density.  For AlN thicknesses lower than this value, the band-bending in the AlN barrier is lower than the bandgap of GaN, $\mathcal{E}_{c}(n-side) > \mathcal{E}_{v}(p-side)$, the physical depletion region width is larger than $t_{AlN}$, and therefore tunneling is prohibited for small reverse bias voltages.  On the other hand, for AlN thicknesses larger than $t_{AlN}^{cr}$, the bands align so as to allow tunneling, but the tunneling probability is lowered significantly.  These qualitative arguments are now provided a quantitative basis.

%=================================================================
%=================================================================

{\em Model:} The tunneling probability $\mathcal{T}_{wkb}$ across a potential barrier $\mathcal{V}(x)$ in the Wentzel-Kramers-Brillouin (WKB) approximation is given by\cite{sze}:
\begin{equation}\label{eq:T}
\mathcal{T}_{wkb} \simeq e^{ -2 \int^{t_{AlN}}_{0} \kappa(x) dx},
\end{equation}
where the spatial variation of the imaginary wavevector $\kappa(x)$ is given by:
\begin{equation}\label{eq:k}
\kappa (x) = \sqrt{\frac{ 2m^{*} }{ \hbar^{2} } \mathcal{V}(x) } = \sqrt{\frac{ 2m^{\star} }{ \hbar^{2} }( \mathcal{E}_{g} + \Delta \mathcal{E}_{c} - q\mathcal{F}_{\pi}x)},
\end{equation}
as can be seen from the barrier potential $\mathcal{V}(x)$ in Fig \ref{fig4}.  Here, $m^{\star}$ is a reduced tunneling effective mass that is related to the curvature of the imaginary part of the bandstructure inside the bandgap.  Defining constants $\kappa_{0} = \sqrt{2 m^{\star} (\mathcal{E}_{g} + \Delta \mathcal{E}_{c})/\hbar^{2} }$ and $t_{0} = (\mathcal{E}_{g} + \Delta \mathcal{E}_{c})/ q \mathcal{F}_{\pi}$, the WKB tunneling probability is given by $\mathcal{T}_{wkb} \simeq \exp{[ -  2/3 \times \kappa_{0} t_{0} [1- (1 - t_{AlN}/t_{0})^{3/2} ] ] } $.

The tunneling current density that flows through the junction is then found by adding the contributions from carriers at various energies\cite{sze}
\begin{equation}\label{eq:I}
J_{T} \propto \int_{0}^{qV} [ f_{v}^{p}( \mathcal{E} ) - f_{c}^{n}(\mathcal{E})] \rho_{v}^{p}(\mathcal{E})  \rho_{c}^{n}(\mathcal{E})  \mathcal{T}_{wkb} d\mathcal{E}.
\label{jtunnel}
\end{equation}
Here $f_{..}$ are the Fermi-Dirac electron occupation functions.  The subscripts $c$ and $v$ stand for the conduction and valence bands respectively, and the superscripts indicate the $n$ and $p$ sides of the tunnel junction.  The density of states (DOS) in the conduction and valence bands are denoted by $\rho_{c}$ and $\rho_{v}$ respectively.  Without loss of generality, if we assume that the applied bias appears on the n-side, then electrons at the valence-band edge of the p-side will be able to tunnel elastically to states in the conduction band on the n-side at an energy $\delta \mathcal{E}$ above the band edge, as indicated in Fig \ref{fig4}.  Since the conduction band in GaN is highly parabolic, the DOS varies as $\rho_{c}^{n} \sim \sqrt{\mathcal{E}}$.  Therefore, the integrand in equation \ref{jtunnel} depends on the thickness of the AlN barrier as $\sqrt{\delta \mathcal{E}} \times \mathcal{T}_{wkb}$.  Thus, the tunneling current density is proportional to
\begin{equation}
J_{T} \propto \sqrt{qV + q \mathcal{F}_{\pi} t_{AlN} - \mathcal{E}_{g} }  \cdot e^{- \frac{2}{3} \kappa_{0} t_{0} [1- (1 - \frac{t_{AlN}}{t_{0}} )^{\frac{3}{2}} ] } . 
\label{jtunneldep} 
\end{equation}
This functional form is plotted against $t_{AlN}$ in Fig \ref{fig4}, and is in excellent agreement with the trend of tunnel current densities measured experimentally.  For the thickest AlN barrier, the tunneling current is higher than expected - this can be attributed to leakage through defects that form due to strain relaxation, as has been reported in an earlier work\cite{caoyu07apl}.

\begin{figure}
\includegraphics[width=3.5in]{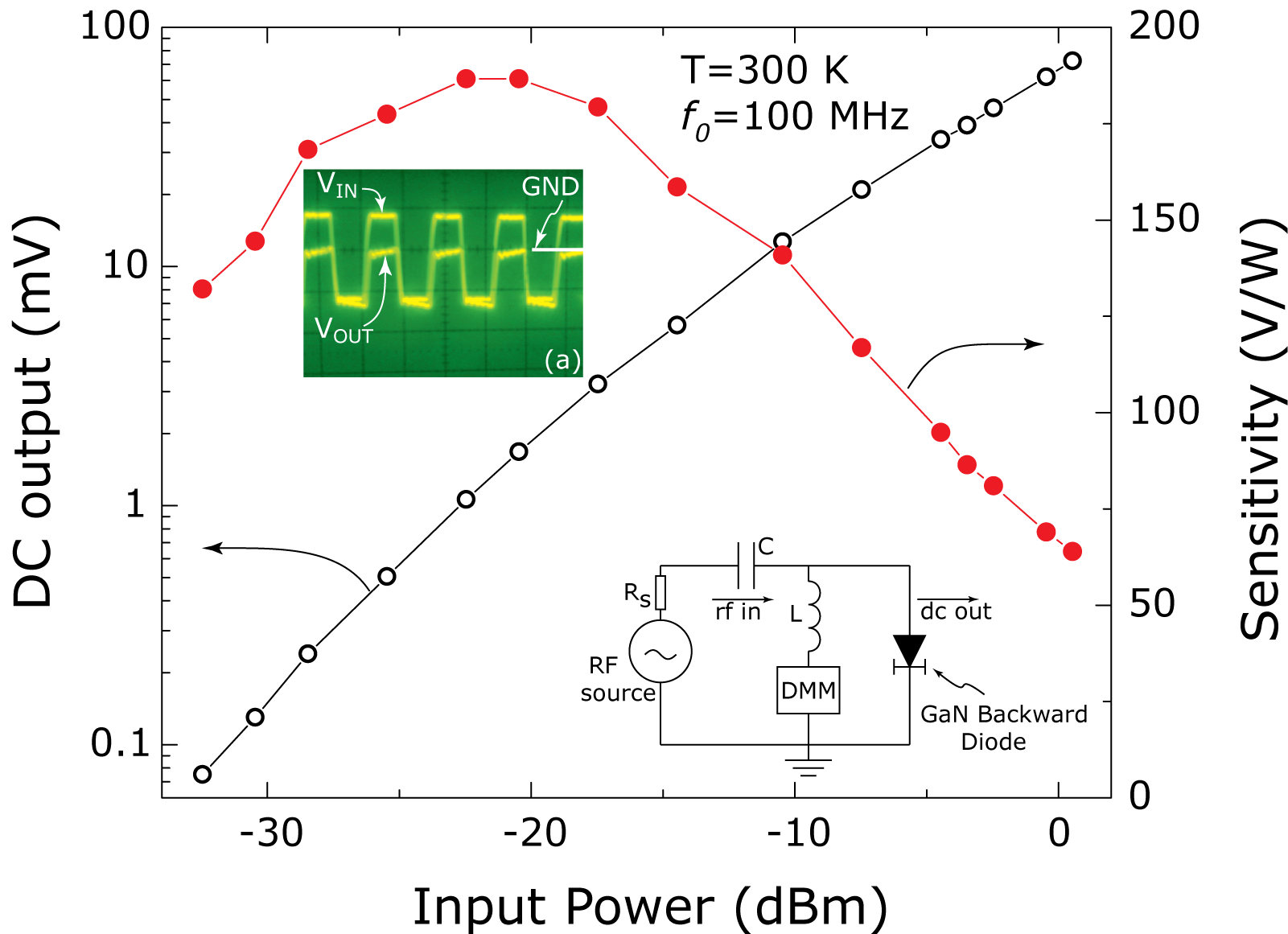}\\
\caption{Measured sensitivity and detector voltage of a large area 56x56 $\mu$m$^{2}$ area backward diode against input power at a RF frequency of 100 MHz.}
\label{fig5}
\vspace{-4ex}
\end{figure}

{\em Rectification, Backward Diodes:}  Backward-diode behavior of the tunnel junctions enable the rectification of oscillatory signals of small voltages.  The ability to perform this operation around zero DC bias makes such devices attractive for mixing, passive mm-wave imaging, and radiometric temperature measurement without the need for external biasing circuitry.  The $t_{AlN} = 2.8$ nm GaN/AlN/GaN tunnel junction was first tested for this property.  Fig \ref{fig5} inset (a) shows that the output of the diode cuts off the positive half of a 0.2 Volt oscillating square-wave input voltage cycle (period: 1 $\mu$s, zero DC voltage), as expected.  The zero-bias curvature, as discussed earlier $\gamma \sim 21 $ V$^{-1}$ is lower than the $\gamma \sim q/kT \sim 38.5 $ V$^{-1}$ `thermal' limit for Schottky diodes.  The curvature in backward diodes is not limited by thermionic emission, and can be improved beyond this classical limit in the future by shrinking the bandgaps and optimizing the doping and the polarization charge at the heterojunction, as has been done for other group-IV and III-V semiconductor materials (for example, see \cite{su08edl} and the references cited therein).   

A 56x56 $\mu$m$^{2}$ area device with $t_{AlN} = 2.8$ nm was then connected to a circuit as shown in Fig \ref{fig5} (inset) to test the rectifying behaviour over a wide range of input voltages.  The sensitivity was measured on-wafer, using a RF power source and measuring the DC return with a digital multimeter (DMM) connected to the device with a bias-T.  The input power was varied from -32 dBm to 0 dBm, at an RF frequency of 100 MHz.  Figure \ref{fig5} shows the measured detector voltage and sensitivity versus the input power.  The detector exhibited square law detection till input power levels of -20 dBm, with the onset of detector compression for higher powers.  This preliminary demonstration shows much promise for future usage of such backward diodes in  higher frequency RF applications with improvements in the heterojunction layer structure and miniaturization of the device geometry.  Owing to the high radiation resistance, and several properties resulting from the large bandgap - high breakdown voltage, optical transparency in the visible, very low interband thermal generation of carriers (and resulting low noise), III-V nitride heterostructure based backward diodes can be useful in the zero-bias detection of very high power RF signals under harsh conditions, where other material systems will be pushed above their physical and thermal performance limits.

%=================================================================
%=================================================================

{\em Discussion \& Conclusions:} Considering the fact that the acceptor doping efficiency in Mg-doped GaN is extremely low (activation energy $E_{A} \sim 200 $ meV) and that the p-type GaN layer is far from being degenerate, it is indeed remarkable that interband tunneling can be observed.  As discussed, the key ingredient that enables this feature is the giant polarization-induced electric field at the heterojunction.  The general paradigm can be extended to other wide-bandgap semiconductor families whose crystal structures allow for high polarization fields (such as the ZnO family).

The tunneling barrier used in this work is AlN, which has the largest bandgap of all III-V and group IV semiconductors.  Though it affords the highest polarization difference between GaN and AlN, the large barrier height also restricts the tunneling current drive.  The tunneling current density scales as $\sim \exp [ - (\mathcal{E}_{g} + \Delta \mathcal{E}_{c})^{3/2} ]$, thereby implying a lower current density than similar devices using narrower-gap semiconductors.  The polarization-induced tunnel diodes studied here for GaN exhibit current densities in the 1-10 A/cm$^2$ range.  While these current densities are low for nitride optical emitters (light-emitting diodes, LASERs), they are sufficiently high for solar cells where typical requirements are in the 10s of mA/cm$^2$.  A major advantage of these tunnel diode structures for solar cells is that the polar barrier that enables tunneling is of wider bandgap and can thus be optically transparent, thereby offering an attractive solution to absorption losses encountered in the tunnel junctions connecting typical multijunction cells.  Furthermore, using narrower-bandgap III-V nitrides (InGaN) in the p- and n-regions and using GaN (or AlGaN) as the tunnel-barrier, the current densities can be enhanced.  Such junctions may then find use in optical emitters as well.  

Going beyond conventional electronic/optical device applications, tunneling measurements have been used as an effective tool for spectroscopy of the electronic structure of semiconductors.  The polarization-induced tunnel junctions can be used for measuring the electronic bandstructure of narrower-gap nitrides (such as InN).  Finally, they offer a method to tunnel-inject spin-polarized carriers and take advantage of the large spin-lifetimes of the wide-bandgap nitrides.

The authors acknowledge useful discussions with S. Koswatta, A. Seabaugh, M. Grundmann and U. K. Mishra.  This work was performed with financial support from the Center for the Engineering of Oxide/Nitride Structures (EONS) and the Office of Naval Research (Dr. P. Maki).

%=================================================================
%CONCLUSIONS & FUTURE WORK

%----------------------------------------------------------------
%Figure Template
%\begin{figure}[h]
%\begin{center}
%\leavevmode \epsfxsize=3in \epsfbox{} \caption{**.}
%\end{center}
%\end{figure}
%Figure
%----------------------------------------------------------------

\bibliographystyle{nature}

\begin{thebibliography}{1}

\bibitem{Esaki1} L. Esaki, Phys. Rev. {\bf109} 603 (1958).

\bibitem{sze} S. M. Sze, {\em Physics of Semiconductor Devices}, 2nd Ed., (Wiley, New York, 1981), Chapter 9.

\bibitem{Persson} D.R. Persson, Radio and Electronic Engineer. {\bf 31} 241 (1966).
    
\bibitem{Chang1} K.K.N. Chang, Proc. of the IRE., {\bf 47} 1268 (1959).

\bibitem{Sterzer} F Sterzer and A Presser, Proc. of the IRE. {\bf 49} 1318 (1961).

\bibitem{Montgomery} M Montgomery, Proc. of the IRE. {\bf 49} 826 (1961).

\bibitem{Miller} D.L. Miller, S.W. Zehr, and J. S. Harris Jr., J. Appl. Phys., {\bf 53} 744 (1982).

\bibitem{DeSalvo} G. C. DeSalvo, J. Appl. Phys. {\bf 74} 4207 (1993).

\bibitem{Bernardini} F. Bernardini, V. Fiorentini, and D. Vanderbilt, Phys. Rev B., {\bf 56}, R10024 (1997).

\bibitem{djPolBook} C. Wood and D. Jena, {\em Polarization Effects in Semiconductors: From Ab Initio Theory to Device Applications}, 1st Ed., (Springer, Berlin, 2007), Chapter 2.

\bibitem{Grundmann} M.J. Grundmann and U.K. Mishra, Phys. Stat. Sol. (c). {\bf 4}, 2830 (2007).

%\bibitem{Sheu} J. K. Sheu, Y. K. Su, G. C. Chi, P. L. Koh, M. J. Jou, C. M. Chang, C. C. Liu, and W. C. Hung {High-transparency \textsc{N}i/\textsc{A}u ohmic contact to p-type \textsc{G}a\textsc{N}.} Appl. Phys. Lett. 74(\textbf{16}),2340 (1999).

\bibitem{caoyu07apl} Y. Cao and D. Jena, Appl. Phys. Lett., {\bf 90} 182112 (2007).

\bibitem{su08edl} N. Su, R. Rajavel, P. Deelman, J. N. Schulman, and P. Fay, IEEE Electron Device Lett., {\bf 29} 536 (2008).

\end{thebibliography}

\end{document}